\documentstyle[preprint,pra,aps]{revtex}
\begin{document}
\draft
\title
{
Quantum state transformation by dispersive and absorbing
four-port devices
}
\author{L. Kn\"oll, S. Scheel, E. Schmidt, and D.-G. Welsch}
\address{Friedrich-Schiller-Universit\"{a}t Jena,
Theoretisch-Physikalisches Institut
\\
Max-Wien-Platz 1, D-07743 Jena, Germany} 
\date{23. 06. 1998} 
\maketitle
\begin{abstract}
The recently derived input--output relations for the
radiation field at a dispersive and absorbing four-port device 
[T. Gruner and D.-G. Welsch, Phys. Rev. A {\bf 54}, 1661 (1996)]
are used to derive the unitary transformation that relates  
the output quantum state to the input quantum state, including
radiation and matter and without placing frequency restrictions.      
It is shown that for each frequency the transformation can be regarded
as a well-behaved SU(4) group transformation that can be decomposed 
into a product of U(2) and SU(2) group transformations. Each of them 
may be thought of as being realized by a particular lossless four-port 
device. If for narrow-bandwidth radiation far from the medium resonances
the absorption  matrix of the four-port device can be disregarded, the 
well-known SU(2) group transformation for a lossless device is 
recognized. Explicit formulas for the transformation of Fock-states
and coherent states are given.  
\end{abstract}
\pacs{PACS number(s): 42.50.-p, 42.50.Ct, 42.25.Bs, 42.79.-e, 
}

\narrowtext
\section{Introduction}
\label{intro}

Four-port devices such as beam splitters are indispensable to 
optical investigation, and a number of fundamental experiments
in quantum optics necessarily require the use of them.
The quantum theory of dispersionless and nonabsorbing
beam splitters has been well established 
\cite{Yurke-McCall-Klauder-1986,Prasad-Scully-Martienssen-1987,%
Ou-Hong-Mandel-1987,Fearn-Loudon-1987,Campos-Saleh-Teich-1989,%
Leonhardt-1993,Luis-Sanchez-Soto-1995}.
A beam splitter can be realized by a multislab dielectric
plate, which is a dispersive and absorbing device in general.
Even if the effects of dispersion and absorption 
(in a chosen frequency interval) are small, their 
influence on nonclassical radiation should be considered 
carefully. On the other hand, in practice multislab
dielectric configurations with strongly varying
dispersive and absorptive properties, e.g., near
optical band gaps, have been of increasing interest,
and a description of their action in the quantum domain 
is desired. 

To give a quantum theory of a dispersive and absorbing
(linear) four-port device, a Kramers--Kronig consistent quantization scheme 
of the electromagnetic field in dispersive and absorbing
inhomogeneous media is required
\cite{Gruner-Welsch-1995,Matloob-Loudon-Barnett-Jeffers-1995,%
Matloob-Loudon-1996,Ho-Knoell-Welsch-1998,Scheel-Knoell-Welsch-1998}.
In particular, quantization of the radiation field within the framework 
of the phenomenological Maxwell theory (with given complex permittivity
in the frequency domain) can be performed using an
expansion of the electromagnetic field operators in terms of
the Green function of the classical problem and an appropriately
chosen infinite set of bosonic basic fields \cite{Gruner-Welsch-1995}.
This quantization scheme, which may be regarded as a generalization of the
familiar concepts of mode expansion, applies to
any inhomogeneous dielectric matter and is consistent
with both the Kramers--Kronig relations and the canonical
(equal-time) field commutation relations in QED
\cite{Ho-Knoell-Welsch-1998,Scheel-Knoell-Welsch-1998}.

The formalism has been used in order to derive input--output
relations for radiation at a dispersive and absorbing (multilayer)
dielectric plate and to express the moments and correlations of
the outgoing fields in terms of those of the incoming fields and the
(initial) dielectric-matter excitations
\cite{Matloob-Loudon-Barnett-Jeffers-1995,%
Matloob-Loudon-1996,Gruner-Welsch-1996,Barnett-Gilson-Huttner-Imoto-1996}.
Such a (multilayer) dielectric plate may serve as a model for a
number of four-port devices, such as beam splitters, mirrors,
thin films, interferometers, and optical fibers.
The results have been used for studying low-order
correlations in two-photon interference effects
\cite{Gruner-Welsch-1996,Gruner-Welsch-1997,%
Barnett-Jeffers-Gatti-Loudon-1998}.

In this paper we extend the input-output relations for the
radiation field at a dispersive and absorbing four-port device
to the complete SU(4) group transformations for radiation and matter,
and present closed formulas for the transformation of the quantum
state as a whole. It is worth noting that the theory applies to
optical fields at arbitrary frequencies and bandwidths. In particular
for narrow-bandwidth light in which the frequencies are far from medium
resonances so that absorption may be disregarded, the well-known
results of SU(2) symmetry are recognized. In the general case of
nonvanishing absorption, for each frequency 
the SU(4) group transformation can be given 
by a product of eight U(2) and SU(2) group
transformations, which correspond to an equivalent network
of eight lossless four-port devices for radiation and matter.

The paper is organized as follows. In Sec.~\ref{inout} the underlying
theory is outlined and the basic input-output relations are given.
The problem of quantum-state transformation is studied in
Sec.~\ref{state} and closed solutions are presented. To illustrate
the theory, explicit transformation rules for Fock states and
coherent states are presented. A summary and some conclusions are
given in Sec.~\ref{sum}.


\section{Basic equations}
\label{inout}

Let us consider two light beams (of fixed 
polarization) that propagate along the (positive) $x_1$ and $x_2$ axes
and impinge on a dispersive and absorbing four-port device 
that gives rise to two outgoing beams propagating along 
the (positive) $y_1$ and $y_2$ axes. Following \cite{Gruner-Welsch-1996},
the operator of the vector potential in each of the four channels 
of the device can be given by
\begin{eqnarray}
\lefteqn{
\hat{A}_j(z_j) 
= \!\!\int_0^{\infty} \!\!\!{\rm d} \omega \,
\Bigg[
\sqrt{\frac{\hbar\beta_j(\omega)}
{4 \pi c \omega \epsilon_0 n_j^2(\omega){\cal A}}}
}
\nonumber \\ && \hspace{17ex} \times
\, e^{i\beta_j(\omega) \omega z_j/c} \hat{c}(z_j,\omega)
\!+ \!{\rm H.c.} 
\Bigg]
\label{2.1}
\end{eqnarray}
($j$ $\!=$ $\!1,2$), where 
\begin{equation}
n_j(\omega) = \sqrt{\epsilon_j(\omega)}
= \beta_j(\omega) + i \, \gamma_j(\omega)
\label{2.2}
\end{equation}
is the complex refractive index of the adjacent medium on the $j$th side 
of the device (${\cal A}$, plane area of the beam). 
In Eq.~(\ref{2.1}), $\hat{c}_j(z_j,\omega)$ stands for the amplitude operators
$\hat{a}_j(x_j,\omega)$ and $\hat{b}_j(y_j,\omega)$, respectively,
of the incoming and outgoing damped waves at frequency $\omega$. The 
input-output relations for the amplitude operators can be derived to be  
\begin{equation}
\hat{b}_j(\bar{y}_j,\omega)
= \sum_{j'=1}^2 T_{jj'}(\omega) \hat{a}_{j'}(\bar{x}_{j'},\omega)
+ \sum_{j'=1}^2 A_{jj'}(\omega) \hat{g}_{j'}(\omega)
\label{2.3}
\end{equation}
where it is assumed that the incoming beams enter the device
at $x_j$ $\!=$ $\!\bar{x}_j$ and the outgoing beams leave the
device at $y_j$ $\!=$ $\!\bar{y}_j$. The operators $\hat{g}_j(\omega)$ 
play the role of operator noise sources and describe  
device excitations. The $2\times 2$ matrices 
$T_{jj'}(\omega)$ and $A_{jj'}(\omega)$
are the characteristic transformation and absorption matrices
of the device. Whereas the $T_{jj'}$ matrix describes the 
effects of reflection and transmission, the $A_{jj'}$ 
matrix results from the losses inside the device (for $T_{jj'}(\omega)$
and $A_{jj'}(\omega)$ of a multilayer dielectric slab,
see \cite{Gruner-Welsch-1996}). Finally, the commutation rules for 
the amplitude operators of the incoming waves and the operators 
of device excitations are 
\begin{eqnarray}
\lefteqn{
\big[ \hat{a}_{j}(x_{j},\omega), \hat{a}_{j'}^{\dagger}(x_{j'}',\omega')\big]  
}
\nonumber\\&&\hspace{5ex}
= \delta_{jj'} \delta(\omega-\omega')
e^{- \gamma_j(\omega)\,\omega |x_j-x_{j'}'|/c} ,
\label{2.4}
\end{eqnarray}
\begin{equation}
\big[\hat{g}_{j}(\omega),\hat{g}_{j'}^{\dagger}(\omega')\big] 
= \delta_{jj'} \delta(\omega - \omega') ,
\label{2.5}
\end{equation}
\begin{equation}
\big[\hat{a}_{j}(x_j,\omega),\hat{g}_{j'}(\omega')^{\dagger} \big] = 0 
\label{2.6}
\end{equation}
($x_j$ $\!\ge$ $\bar{x}_j$). Since the dependence on space of the 
amplitude operators outside the device is governed by quantum 
Langevin equations, the input-output relations (\ref{2.3}) together 
with the commutation relations (\ref{2.4}) -- (\ref{2.6}) fully
determine the action of the device. The commutation relations 
(\ref{2.4}) -- (\ref{2.6}) reveal that the amplitude operators of 
the incoming waves at the entrance plane, 
$\hat{a}_{j}(\omega)$ $\!\equiv$ $\!\hat{a}_{j}(\bar{x}_j,\omega)$, 
and the operators of the device excitations, $\hat{g}_{j}(\omega)$
are independent bosonic operators. The amplitude operators 
of the outgoing waves, 
$\hat{b}_{j}(\omega)$ $\!\equiv$ $\!\hat{b}_{j}(\bar{y}_j,\omega)$, 
do not satisfy bosonic commutation relations in general. For given matrices 
$T_{jj'}(\omega)$ and $A_{jj'}(\omega)$,
their commutation relations can be derived straightforwardly, applying 
Eq.~(\ref{2.3}) and using Eqs.~(\ref{2.4}) -- (\ref{2.6}). 

Let us consider the case when
the device is surrounded by vacuum [$n_j(\omega)$ $\!\to$ $\!1$].
In this case the amplitude operators of the 
incoming and outgoing waves become independent of space 
and reduce to ordinary bosonic operators. In particular
it can be shown that the matrix relation
\begin{equation}
\sum_{k=1}^2 T_{jk}(\omega) T_{j'k}^\ast(\omega) 
+ \sum_{k=1}^2 A_{jk}(\omega) A_{j'k}^\ast(\omega) = \delta_{jj'}
\label{2.8}
\end{equation}
is valid, which implies the bosonic commutation relation
\begin{equation}
\big[ \hat{b}_j(\omega), \hat{b}_{j'}^{\dagger}(\omega')\big]  
=  \delta_{jj'} \delta(\omega-\omega') .
\label{2.9}
\end{equation}
Note that the relation (\ref{2.8}) reflects the fact that
when the device is embedded in vacuum, the sum of the 
probabilities for reflection, transmission, and absorption of a
photon is equal to one. 
When the device is embedded in a medium, the matrix relation
(\ref{2.8}) and the bosonic commutation relation
(\ref{2.9}) are not valid in general. From Eqs.~(\ref{2.3})
-- (\ref{2.6}) it can be seen that a unitary transformation
\begin{equation} 
\label{2.9a}
\hat{b}'_i(\omega) = \sum_{k=1}^{2} X_{ik}(\omega) \hat{b}_k(\omega)
\end{equation}
[$(X^{-1})_{ik}$ $\!=$ $\!X_{ki}^\ast$]
can be introduced such that
\begin{equation}
\left[ \lambda^{-\frac{1}{2}}_j(\omega)\hat{b}'_j(\omega), 
\lambda^{-\frac{1}{2}}_{j'}(\omega)\hat{b}'_{j'}{^{\dagger}}(\omega')\right]  
=  \delta_{jj'} \delta(\omega-\omega') 
\label{2.9b}
\end{equation}
($\lambda_j$ $\!>$ $\!0$). Hence the transformed and scaled
operators $\lambda^{-\frac{1}{2}}_j(\omega)\hat{b}'_j(\omega)$ 
are bosonic operators and the corresponding (scaled and
transformed) transformation and absorption matrices satisfy the
condition (\ref{2.8}). 

Without loss of generality we can therefore restrict our attention
to a bosonic system and assume that the matrix relation (\ref{2.8})
is valid. For notational reasons it is convenient to  
introduce the definitions
\begin{equation}
\label{2.10}
\hat{\bf a}(\omega)
=\left(\begin{array}{c}\hat{a}_1(\omega)
\\\hat{a}_2(\omega)\end{array}\right), 
\end{equation}
\begin{equation}
\label{2.11}
\hat{\bf g}(\omega)
=\left(\begin{array}{c}\hat{g}_1(\omega)\\\hat{g}_2(\omega)\end{array}\right),
\end{equation}
\begin{equation}
\label{2.12}
\hat{\bf b}(\omega)
=\left(\begin{array}{c}\hat{b}_1(\omega)\\\hat{b}_2(\omega)\end{array}\right)
\end{equation}
and
\begin{equation}
\label{2.13}
{\bf T}(\omega)
=\left(\begin{array}{cc}T_{11}(\omega)&T_{12}(\omega)\\
T_{21(\omega)}&T_{22}(\omega)\end{array}\right),
\end{equation}
\begin{equation}
\label{2.14}
{\bf A}(\omega)
=\left(\begin{array}{cc}A_{11}(\omega)&A_{12}(\omega)\\
A_{21}(\omega)&A_{22}(\omega)\end{array}\right).
\end{equation}
The input-output relations for radiation at a general four-port device
can then be given in the compact form of
\begin{equation}
\label{2.15} 
\hat{\bf b}(\omega) = {\bf T}(\omega) \hat{\bf a}(\omega) 
+ {\bf A}(\omega) \hat{\bf g}(\omega) ,
\end{equation}
with 
\begin{equation}
\label{2.16}
{\bf T}(\omega){\bf T}^+(\omega) 
+ {\bf A}(\omega){\bf A}^+(\omega) ={\bf I} .
\end{equation}


\section{Quantum state transformation}
\label{state}

The operator input-output relation (\ref{2.15}) enables one to
calculate arbitrary correlations of the outgoing beams
from the correlations of the incoming beams and
the device excitations \cite{Gruner-Welsch-1996}. To obtain 
the quantum state of the outgoing beams as a whole,
the question arises which quantum state transformation
corresponds to the operator input-output relation.
Let us assume that the incoming fields and the device are prepared in
a quantum state described by the density operator $\hat{\varrho}_{\rm in}$ 
and that for any frequency the input-output relation (\ref{2.15}) 
corresponds to the unitary operator transformation
\begin{equation}
\hat{\bf b}(\omega) 
= \hat{U}^\dagger \hat{\bf a}(\omega) \hat{U},
\quad \hat{U}^\dagger = \hat{U}^{-1} .
\label{2.17}
\end{equation}
The effect of the device can equivalently be described by
leaving the photonic operators $\hat{a}_j(\omega)$ unchanged
but transforming the input-state density operator
$\hat{\varrho}_{\rm in}$ to obtain the 
output-state density operator $\hat{\varrho}_{\rm out}$ as
\begin{equation}
\hat{\varrho}_{\rm out}
= \hat{U} \hat{\varrho}_{\rm in} \hat{U}^\dagger .
\label{2.18}
\end{equation}

   
\subsection{Lossless device}
\label{beam-splitter}

Let us first restrict our attention to a field in a sufficiently
small frequency interval of width $\Delta\omega$ in which absorption
may be disregarded. For this frequency window the four-port device
can be regarded as being lossless, and Eqs.~(\ref{2.15}) and 
(\ref{2.16}) reduce to
\begin{equation}
\label{2.19} 
\hat{\bf b}(\omega) = {\bf T}(\omega) \hat{\bf a}(\omega) ,
\end{equation}
\begin{equation}
\label{2.20}
{\bf T}(\omega){\bf T}^+(\omega) = {\bf I} .
\end{equation}
Here, the elements of the U(2) matrix ${\bf T}(\omega)$
are usually given by
\begin{eqnarray}
T_{jj'}(\omega) = 
\left(\begin{array}{rr} t(\omega) & r(\omega) \\
- r^\ast(\omega) & t^\ast(\omega)
\end{array}\right) e^{i\varphi(\omega)} ,
\label{2.21}
\end{eqnarray}
where $t(\omega)$ and $r(\omega)$, respectively, correspond to the 
complex transmittance and reflectance of the device at frequency
$\omega$,
\begin{equation}
\label{2.22}
t(\omega) = \cos\theta(\omega) \, e^{i\alpha(\omega)}, 
\quad 
r = \sin\theta(\omega) \, e^{i\beta(\omega)}.
\end{equation}
When the phase shift $\varphi(\omega)$ can be disregarded, then
the U(2) group transformation reduces to an SU(2) group
transformation. Note that the phase shift can always be
included in the input operators by replacing $\hat{\bf a}(\omega)$
with $\hat{\bf a}(\omega)e^{i\varphi(\omega)}$, so that
${\bf T}(\omega)$ becomes an SU(2) group matrix.
The unitary exponential operator $\hat{U}$ in Eqs.~(\ref{2.17})
and (\ref{2.18}) can easily be found by extending the formalism 
of lossless beam-splitter transformation 
\cite{Yurke-McCall-Klauder-1986,Campos-Saleh-Teich-1989} to 
multi-mode fields:
\begin{equation}
\label{2.24}
\hat{U} = \exp\!\left[-i \int_{\Delta\omega} {\rm d}\omega \,
\big(\hat{\bf a}^\dagger(\omega)\big)^T 
{\bf V}(\omega) \hat{\bf a}(\omega)\right]
\end{equation}
(the superscript $T$ introduces transposition), where
the $2\times 2$ Hermitian matrix ${\bf V}(\omega)$ is related to the 
SU(2) matrix ${\bf T}(\omega)$ in Eq.~(\ref{2.19}) as 
\begin{equation}
\exp\!\left[-i {\bf V}(\omega) \right] = {\bf T}(\omega).
\label{2.25}
\end{equation}
The operator $\hat{U}$ can be factored in different ways, e.g., 
\begin{eqnarray}
\label{2.26}
\lefteqn{
\hat{U} = 
\exp\!\left\{i \int_{\Delta\omega} {\rm d}\omega\,
\varphi(\omega) \big[
\hat{a}_1^{\dagger}(\omega) \hat{a}_1(\omega)
+ \hat{a}_2^{\dagger}(\omega) \hat{a}_2(\omega)\big]
\right\}
}
\nonumber \\ && \hspace{10ex}\times\,
\exp\!\left[\int_{\Delta\omega} {\rm d}\omega\,\ln t(\omega)\,
\hat{a}_1^{\dagger}(\omega) \hat{a}_1(\omega)
\right]
\nonumber \\ && \hspace{10ex}\times\,
\exp\!\left[-\int_{\Delta\omega} {\rm d}\omega\,r^\ast(\omega)\,
\hat{a}_2^{\dagger}(\omega)\hat{a}_1(\omega)
\right]
\nonumber \\ && \hspace{10ex}\times\,
\exp\!\left[\int_{\Delta\omega} {\rm d}\omega\,r(\omega)\,
\hat{a}_1^{\dagger}(\omega)\hat{a}_2(\omega)
\right]
\nonumber \\ && \hspace{10ex}\times\,
\exp\!\left[-\int_{\Delta\omega} {\rm d}\omega\,\ln t(\omega)\,
\hat{a}_2^{\dagger}(\omega)\hat{a}_2(\omega)
\right].
\end{eqnarray}


\subsection{Dispersive and absorbing device}
\label{lossy device}

\subsubsection{Transformation law}
\label{general}

In order to apply the input--output relation (\ref{2.15}) [together
with Eq.~(\ref{2.16})], we first extend it
to a U(4) group transformation. For this purpose
we combine the two-dimensional vectors $\hat{\bf a}(\omega)$
and $\hat{\bf g}(\omega)$ to obtain a four-dimensional input vector
\begin{equation}
\label{2.27}
\hat{\mbox{\boldmath $\alpha$}}(\omega)
=\left(\begin{array}{c}{\hat{\bf a}(\omega)}
\\{\hat{\bf g}(\omega)}\end{array}\right) 
\end{equation}
and supply the two-dimensional vector $\hat{\bf b}(\omega)$
with some other two-dimensional vector $\hat{\bf h}(\omega)$ 
to obtain a four-dimensional output vector
\begin{equation}
\label{2.27a}
\hat{\mbox{\boldmath $\beta$}}(\omega)
=\left(\begin{array}{c}{\hat{\bf b}(\omega)}
\\{\hat{\bf h}(\omega)}\end{array}\right) .
\end{equation}
Now we relate the four-dimensional vector 
$\hat{\mbox{\boldmath $\beta$}}(\omega)$ to the four-dimensional 
vector $\hat{\mbox{\boldmath $\alpha$}}(\omega)$ as
\begin{equation}
\label{2.28} 
\hat{\mbox{\boldmath $\beta$}}(\omega) 
= \mbox{\boldmath $\Lambda$}(\omega)
\hat{\mbox{\boldmath $\alpha$}}(\omega) , 
\end{equation}
\begin{equation}
\label{2.29} 
\mbox{\boldmath $\Lambda$}(\omega)
\mbox{\boldmath $\Lambda$}^+(\omega) 
= \mbox{\boldmath $I$} , 
\end{equation}
where the $4\times 4$ unitary matrix $\mbox{\boldmath $\Lambda$}(\omega)$
is chosen such that the input--output relation (\ref{2.15}) between 
$\hat{\bf b}(\omega)$ and $\hat{\bf a}(\omega)$ is preserved.
The matrix $\mbox{\boldmath $\Lambda$}(\omega)$ can be expressed in 
terms of $2\times 2$ matrices as (App.~\ref{U4})  
\begin{equation}
\label{2.30}
\mbox{\boldmath $\Lambda$}(\omega)
=\left(\begin{array}{cc}{\bf T}(\omega)
&{\bf A}(\omega)\\[1ex]
-{\bf S}(\omega){\bf C}^{-1}(\omega){\bf T}(\omega)
&{\bf C}(\omega){\bf S}^{-1}(\omega){\bf A}(\omega)\end{array}\right),
\end{equation}
where 
\begin{equation}
\label{2.35}
{\bf C}(\omega) = \sqrt{{\bf T}(\omega){\bf T}^+(\omega)} 
\end{equation}
and
\begin{equation}
\label{2.35a}
{\bf S}(\omega) = \sqrt{{\bf A}(\omega){\bf A}^+(\omega)}
\end{equation}
are commuting positive Hermitian matrices, and
\begin{equation}
\label{2.36}
{\bf C}(\omega)^2 +  {\bf S}(\omega)^2 = {\bf I} .
\end{equation}
In Eq.~(\ref{2.30}) the unitary matrix 
${\bf D(\omega)}$ that appears in Eq.~(\ref{B13}) in App.~\ref{U4} 
has been omitted, since it corresponds to an irrelevant change of 
the device variables $\hat{\bf h}(\omega)$, as it can be seen from the
second line in the large brackets in Eq.~(\ref{2.30}). 
Note that after separation of phase factors
$e^{i\varphi(\omega)}$ and $e^{i\psi(\omega)}$, respectively,
from the matrices ${\bf T}(\omega)$ and ${\bf A}(\omega)$ and
inclusion of them in the operators $\hat{\bf a}(\omega)$ and
$\hat{\bf g}(\omega)$ the matrix $\mbox{\boldmath $\Lambda$}(\omega)$
can be regarded as an SU(4) matrix. 

The input-output relation (\ref{2.28}) can be
expressed in terms of a unitary operator transformation
\begin{equation}
\label{2.37} 
\hat{\mbox{\boldmath $\beta$}}(\omega) = 
\hat{U}^\dagger 
\hat{\mbox{\boldmath $\alpha$}}(\omega) \hat{U},
\end{equation}
where the unitary operator $\hat{U}$ is given by 
\begin{equation}
\label{2.38}
\hat{U} = \exp\!\left[-i \int_0^\infty {\rm d}\omega \,
\big(\hat{\mbox{\boldmath $\alpha$}}^\dagger(\omega)\big)^T 
\mbox{\boldmath $\Phi$}(\omega) 
\hat{\mbox{\boldmath $\alpha$}}(\omega)\right]. 
\end{equation}
Here, $\mbox{\boldmath $\Phi$}(\omega)$ is a $4\times 4$ Hermitian matrix 
which is related to the SU(4) matrix $\mbox{\boldmath $\Lambda$}(\omega)$ by
\begin{equation}
\exp\!\left[-i \mbox{\boldmath $\Phi$}(\omega) \right] 
= \mbox{\boldmath $\Lambda$}(\omega) .
\label{2.39}
\end{equation}
Note that for narrow-bandwidth radiation far from medium 
resonances the $\omega$ integral in Eq.~(\ref{2.38})
can be restricted to a small interval in which absorption may
be disregarded, ${\bf A}(\omega)$ $\!\approx$ $\!{\bf 0}$,
and hence 
\begin{equation}
\label{2.40}
\mbox{\boldmath $\Lambda$}(\omega) 
\approx \left(\begin{array}{cc}{\bf T}(\omega)&{\bf 0}\\
{\bf 0}&{\bf I}\end{array}\right),
\end{equation}
\begin{equation}
\label{2.41}
\mbox{\boldmath $\Phi$}(\omega) 
\approx \left(\begin{array}{cc}{\bf V}(\omega)&{\bf 0}\\
{\bf 0}&{\bf 0}\end{array}\right) .
\end{equation}
In this case Eq.~(\ref{2.38}) approximately reduces to
Eq.~(\ref{2.24}) and the SU(2) group transformation for a 
lossless device is recognized. 

Obviously, the first and the second line in the vector equation 
(\ref{2.37}) correspond to the input--output relation (\ref{2.17}). 
Application of Eq.~(\ref{2.17}) then yields the 
output quantum state $\hat{\varrho}_{\rm out}$, from which the
quantum state of the outgoing radiation field, 
$\hat{\varrho}_{\rm out}^{({\rm F})}$, can be derived,
\begin{equation}
\label{2.43}
\hat{\varrho}_{\rm out}^{({\rm F})}
= {\rm Tr}^{({\rm D})}\big\{\hat{\varrho}_{\rm out}\big\}
={\rm Tr}^{({\rm D})}
\big\{\hat{U} \hat{\varrho}_{\rm in} \hat{U}^\dagger \big\} , 
\end{equation}
where ${\rm Tr}^{({\rm D})}$ means the trace with respect to the device. 
The input density operator $\hat{\varrho}_{\rm in}$ is an operator 
functional of $\hat{\mbox{\boldmath $\alpha$}}(\omega)$ and 
$\hat{\mbox{\boldmath $\alpha$}}^\dagger(\omega)$,
\begin{equation}
\label{2.43a}
\hat{\varrho}_{\rm in} = 
\hat{\varrho}_{\rm in}\big[\hat{\mbox{\boldmath $\alpha$}}(\omega),
\hat{\mbox{\boldmath $\alpha$}}^\dagger(\omega)\big] , 
\end{equation}
and hence the transformed density operator 
$\hat{\varrho}_{\rm out}$
can be given by 
\begin{equation}
\label{2.43b}
\hat{\varrho}_{\rm out} =
\hat{\varrho}_{\rm in}\big[
\hat{U}\hat{\mbox{\boldmath $\alpha$}}(\omega)\hat{U}^\dagger,
\hat{U}\hat{\mbox{\boldmath $\alpha$}}^\dagger(\omega)\hat{U}^\dagger\big] . 
\end{equation}
Recalling Eqs.~(\ref{2.28}) and (\ref{2.37}), we see that
\begin{eqnarray}
\label{2.43c}
\hat{U}\hat{\mbox{\boldmath $\alpha$}}(\omega)\hat{U}^\dagger 
= \mbox{\boldmath $\Lambda$}^+(\omega)
\hat{\mbox{\boldmath $\alpha$}}(\omega),
\end{eqnarray}
\begin{eqnarray}
\label{2.43c1}
\hat{U}\hat{\mbox{\boldmath $\alpha$}}^\dagger(\omega)\hat{U}^\dagger 
= \mbox{\boldmath $\Lambda$}^T(\omega)
\hat{\mbox{\boldmath $\alpha$}}^\dagger(\omega) .
\end{eqnarray}
Combining Eqs.~(\ref{2.43}) -- (\ref{2.43c1}), we derive
\begin{equation}
\label{2.43d}
\hat{\varrho}_{\rm out}^{({\rm F})}
= {\rm Tr}^{({\rm D})}\!\left\{\hat{\varrho}_{\rm in} 
\big[\mbox{\boldmath $\Lambda$}^+(\omega)
\hat{\mbox{\boldmath $\alpha$}}(\omega),
\mbox{\boldmath $\Lambda$}^T(\omega)
\hat{\mbox{\boldmath $\alpha$}}^\dagger(\omega)\big]
\right\} . 
\end{equation}


\subsubsection{Relation to U(2) and SU(2) group transformations}
\label{two}

As shown in App.~\ref{app}, the U(4) group transformation defined
by the matrix $\mbox{\boldmath $\Lambda$}$ given in Eq.~(\ref{B13})
is equivalent to five U(2) group transformations. That is to say,
for chosen frequency the action of an absorbing four-port device
formally corresponds to the combined action of five lossless
four-port devices in general (for the factorization of an
U(N) matrix into U(2) matrices, see also 
\cite{Reck-Zeilinger-Bernstein-Bertani-1994}). 
When the irrelevant matrix 
${\bf D}(\omega)$ in Eq.~(\ref{B13}) is set equal to the unit matrix
${\bf I}$, then Eq.~(\ref{B13}) reduces to Eq.~(\ref{2.30}) and
the action of the absorbing device corresponds, for
chosen frequency, reduces to the combined action of eight lossless 
devices in general. In this case the unitary
operator (\ref{2.38}) can be factored into a product of 
unitary operators of the type given in Eq.~(\ref{2.24}) for 
a lossless device,
\begin{equation}
\label{2.44}
\hat{U}[{\bf M};\hat{\bf q}] 
\equiv \exp\!\left[-i \int_0^\infty {\rm d}\omega \,
\big(\hat{\bf q}^\dagger(\omega)\big)^T 
{\bf W}(\omega) \hat{\bf q}(\omega)\right] .
\end{equation}
Here, ${\bf W}(\omega)$ is a $2\times 2$ Hermitian 
matrix that is related to a U(2) group transformation
matrix ${\bf M}(\omega)$ as
\begin{equation}
\label{2.44a}
\exp[-i{\bf W}(\omega)] = {\bf M}(\omega),
\end{equation}
and $\hat{\bf q}(\omega)$ is a vector whose two components are
bosonic operators.
Note that for narrow-bandwidth radiation far from medium resonances
Eq.~(\ref{2.44}) [together with Eq.~(\ref{2.44a})]
corresponds to Eq.~(\ref{2.24}) [together with Eq.~(\ref{2.25})],
with ${\bf M}(\omega)$ $\!=$ $\!{\bf T}(\omega)$,
${\bf W}(\omega)$ $\!=$ $\!{\bf V}(\omega)$,
and $\hat{\bf q}(\omega)$ $\!=$ $\!\hat{\bf a}(\omega)$. 
As shown in App.~\ref{app}, the
unitary operator given in Eq.~(\ref{2.38}),
\begin{equation}
\label{2.44b}
U\equiv
\hat{U}[\mbox{\boldmath $\Lambda$}; 
\hat{\mbox{\boldmath $\alpha$}}]
= \exp\!\left[-i \int_0^\infty {\rm d}\omega \,
\big(\hat{\mbox{\boldmath $\alpha$}}^\dagger\big)^T(\omega)
\mbox{\boldmath $\Phi$}(\omega) 
\hat{\mbox{\boldmath $\alpha$}}(\omega)\right],
\end{equation}
can be decomposed into a product of
operators $U[{\bf M};\hat{\bf q}]$ as follows:
\begin{eqnarray}
\label{2.45}
\lefteqn{
\hat{U}[\mbox{\boldmath $\Lambda$};\hat{\mbox{\boldmath $\alpha$}}]
= \hat{U}[{\bf C}\!+\!i{\bf S};(i\hat{\bf a}\!+\!\hat{\bf g})/\sqrt{2}]
}
\nonumber\\ & & \hspace{2ex}\times 
\hat{U}[{\bf C}\!-\!i{\bf S};(\hat{\bf a}\!+\!i\hat{\bf g})/\sqrt{2}]
\,\hat{U}[{\bf S}^{-1}{\bf A};\hat{\bf g}]
\,\hat{U}[{\bf C}^{-1}{\bf T};\hat{\bf a}] 
\end{eqnarray} 
[cf. Eqs.~(\ref{A07a}), (\ref{A10a}), and (\ref{A17})],
and decomposition of 
$\hat{U}[{\bf C}\!-\!i{\bf S};(\hat{\bf a}\!+\!i\hat{\bf g})/\sqrt{2}]$
and $\hat{U}[{\bf C}\!+\!i{\bf S};(i\hat{\bf a}\!+\!\hat{\bf g})/\sqrt{2}]$
eventually yields 
\begin{eqnarray}
\label{2.45a}
\lefteqn{
\hat{U}[\mbox{\boldmath $\Lambda$}; \hat{\mbox{\boldmath $\alpha$}}]
= 
\hat{U}^\dagger[{\bf P};\hat{\bf d}_2]
\, \hat{U}^\dagger[{\bf P};\hat{\bf d}_1]
}
\nonumber\\ & & \hspace{12ex} \times 
\,\hat{U}[{\bf C}\!+\!i{\bf S};\hat{\bf g}]
\, \hat{U}[{\bf C}\!-\!i{\bf S};\hat{\bf a}]
\nonumber\\ & & \hspace{12ex}\times 
\,\hat{U}[{\bf P};\hat{\bf d}_2] 
\, \hat{U}[{\bf P};\hat{\bf d}_1]
\nonumber\\ & & \hspace{12ex} \times
\,\hat{U}[{\bf S}^{-1}{\bf A};\hat{\bf g}]
\, \hat{U}[{\bf C}^{-1}{\bf T};\hat{\bf a}],
\end{eqnarray}
where 
\begin{eqnarray}
\label{2.45b}
\hat{\bf d}_j(\omega)
= \left(
\begin{array}{c}
\hat{a}_j(\omega) \\
\hat{g}_j(\omega)
\end{array}
\right) 
\end{eqnarray}
($j$ $\!=$ $\!1,2$) and
\begin{eqnarray}
\label{2.45c}
{\bf P} = \frac{1}{\sqrt{2}}
\left(
\begin{array}{cc}1&i\\i&1
\end{array}
\right)
\end{eqnarray}
[cf. Eqs.~(\ref{A17b}) and (\ref{A17c})].
It should be pointed out
that when $\mbox{\boldmath $\Lambda(\omega)$}$ is an
SU(4) group transformation, then the matrices ${\bf P}$,
${\bf S}^{-1}(\omega){\bf A}(\omega)$, and
${\bf C}^{-1}(\omega){\bf T}(\omega)$
correspond to SU(2) group transformations. The matrices
${\bf C}(\omega)$ $\!+$ $\!i{\bf S}(\omega)$
and ${\bf C}(\omega)$ $\!-$ $\!i{\bf S}(\omega)$
correspond to U(2) group transformations in general, i.e., SU(2)
transformations and additional phase shifts.
Needless to say that each of the operators
$\hat{U}[{\bf M};\hat{\bf q}]$ on the right-hand side in 
Eq.~(\ref{2.45a}) can be further factored, e.g., according
to Eq.~(\ref{2.26}).


\subsubsection{Discretization}
\label{disc}

In quantum optics radiation fields are frequently described
in terms of discrete modes. Here we restrict attention to
(quasi-)monochromatic discrete modes. For this purpose
we subdivide the frequency axis into sufficiently small
intervals $\Delta_m$ with midfrequencies $\omega_m$ and
define the bosonic input operators 
\begin{equation}
\label{2.46}
\hat{\mbox{\boldmath $\alpha$}}_m
= \frac{1}{\sqrt{\Delta_m}} \int_{\Delta_m} {\rm d}\omega \,
\hat{\mbox{\boldmath $\alpha$}}(\omega)
\end{equation}
and the bosonic output operators 
$\hat{\mbox{\boldmath $\beta$}}_m$ accordingly.
The operator input--output relation (\ref{2.28}) then reads as
\begin{equation}
\label{2.47} 
\hat{\mbox{\boldmath $\beta$}}_m 
= \mbox{\boldmath $\Lambda$}_m
\hat{\mbox{\boldmath $\alpha$}}_m, 
\end{equation}
[$\mbox{\boldmath $\Lambda$}_m$ $\!=$
$\mbox{\boldmath $\Lambda$}(\omega_m)$],
which can be rewritten as
\begin{equation}
\label{2.51} 
\hat{\mbox{\boldmath $\beta$}}_m 
=  \hat{U}^\dagger \hat{\mbox{\boldmath $\alpha$}}_m \hat{U}
= \hat{U}_m^\dagger \hat{\mbox{\boldmath $\alpha$}}_m \hat{U}_m \,,
\end{equation}
where [in place of (\ref{2.38})] 
\begin{equation}
\label{2.52}
\hat{U} = \prod_m \hat{U}_m \,,
\end{equation}
with
\begin{equation}
\label{2.53}
\hat{U}_m 
= \exp\!\left[-i 
\big(\hat{\mbox{\boldmath $\alpha$}}_m^\dagger\big)^T
\mbox{\boldmath $\Phi$}_m
\hat{\mbox{\boldmath $\alpha$}}_m
\right] , 
\end{equation}
and according to Eq.~(\ref{2.39}),
the $4\times 4$ Hermitian matrix $\mbox{\boldmath $\Phi_m$}$ is
related to the SU(4) matrix $\mbox{\boldmath $\Lambda_m$}$ as
\begin{equation}
\exp\!\left[-i \mbox{\boldmath $\Phi_m$} \right]
= \mbox{\boldmath $\Lambda_m$} \, .
\label{2.54}
\end{equation}
The input density operator $\hat{\varrho}_{\rm in}$ is now an operator
function of $\hat{\mbox{\boldmath $\alpha$}}_m$ and
$\hat{\mbox{\boldmath $\alpha$}}_m^\dagger$, and according
to Eq.~(\ref{2.43d}), the density operator of the outgoing
radiation field can be given by
\begin{equation}
\label{2.55}
\hat{\varrho}_{\rm out}^{({\rm F})}
= {\rm Tr}^{({\rm D})}\!\left\{\hat{\varrho}_{\rm in} 
\big[\mbox{\boldmath $\Lambda$}_m^+
\hat{\mbox{\boldmath $\alpha$}}_m,
\mbox{\boldmath $\Lambda$}^T_m
\hat{\mbox{\boldmath $\alpha$}}_m^\dagger\big]
\right\} . 
\end{equation}
In close analogy to Eqs.~(\ref{2.45}) and (\ref{2.45a}),
each SU(4)-group transformation operator $\hat{U}_m$
can be decomposed into a product of U(2)- and SU(2)-group 
transformation operators $\hat{U}_m[{\bf M}_m;\hat{\bf q}_m]$,
\begin{eqnarray}
\label{2.56}
\lefteqn{
\hat{U}_m[\mbox{\boldmath $\Lambda_m$}; \hat{\mbox{\boldmath $\alpha$}}_m]
= \hat{U}_m[{\bf C}_m\!+\!i{\bf S}_m;(i\hat{\bf a}_m\!
+\!\hat{\bf g}_m)/\sqrt{2}]
}
\nonumber\\ & & \hspace{8ex}\times 
\,\hat{U}_m[{\bf C}_m\!-\!i{\bf S}_m;(\hat{\bf a}_m\!+\!i\hat{\bf g}_m)/\sqrt{2}]
\nonumber\\ & & \hspace{8ex}\times 
\,\hat{U}[{\bf S}_m^{-1}{\bf A}_m;\hat{\bf g}_m]
\,\hat{U}[{\bf C}_m^{-1}{\bf T}_m;\hat{\bf a}_m] 
\nonumber\\[1ex] & & \hspace{13.5ex}  
= 
\hat{U}^\dagger_m[{\bf P};\hat{\bf d}_{m2}]
\, \hat{U}^\dagger_m[{\bf P};\hat{\bf d}_{m1}]
\nonumber\\ & & \hspace{8ex} \times 
\,\hat{U}_m[{\bf C}_m\!+\!i{\bf S}_m;\hat{\bf g}_m]
\, \hat{U}_m[{\bf C}_m\!-\!i{\bf S}_m;\hat{\bf a}_m]
\nonumber\\ & & \hspace{8ex}\times
\,\hat{U}[{\bf P};\hat{\bf d}_{m2}]
\, \hat{U}_m[{\bf P};\hat{\bf d}_{m1}]
\nonumber\\ & & \hspace{8ex} \times
\,\hat{U}_m[{\bf S}_m^{-1}{\bf A}_m;\hat{\bf g}_m]
\, \hat{U}_m[{\bf C}_m^{-1}{\bf T}_m;\hat{\bf a}_m]\,.
\end{eqnarray}
Here $\hat{U}_m[{\bf M}_m;\hat{\bf q}_m]$ is given by  
\begin{equation}
\label{2.56a}
\hat{U}_m[{\bf M}_m;\hat{\bf q}_m] 
\equiv \exp\!\left[-i 
\big(\hat{\bf q}_m^\dagger\big)^T 
{\bf W}_m \hat{\bf q}_m\right],
\end{equation}
where 
\begin{equation}
\label{2.56b}
\exp[-i{\bf W}_m] = {\bf M}_m
\end{equation}
[cf. Eqs.~(\ref{2.44}) and (\ref{2.44a})]. Recalling
the definition of discrete operators, Eq.~(\ref{2.46}), 
application of Eq.~(\ref{2.26}) to $\hat{U}_m[{\bf M}_m;\hat{\bf q}_m]$
for further factorization is straightforward.


\subsubsection{Transformation of Fock-states and coherent states}
\label{fock}

To illustrate the theory, let us consider the transformation
of Fock states and coherent states as fundamental basis states
for quantum state representation. For the sake of transparency,
we will restrict attention to single-mode states at some
chosen frequency, so that the subscript $m$ can be omitted.
The results can easily be extended to multimode fields by
taking the direct product of single-mode states.
Let be
\begin{eqnarray}
\label{2.61a}
\hat{\varrho}_{\rm in}
= |\psi_{\rm in}\rangle\langle\psi_{\rm in}|,
\end{eqnarray}
\begin{eqnarray}
\label{2.61b}
|\psi_{\rm in}\rangle
= |n_1;n_2,n_3;n_4\rangle
= \prod_{\nu=1}^4
\frac{\hat{\alpha}_\nu^\dagger {^{n_\nu}}}{\sqrt{n_\nu!}}
\,|0\rangle ,
\end{eqnarray}
the density operator of the system in the case when $n_1$
and $n_2$ photons impinge on the device that is excited
in Fock states with $n_3$ and $n_4$ quanta. From
Eq.~(\ref{2.55}) we then obtain
\begin{equation}
\label{2.61c}
\hat{\varrho}_{\rm out}^{({\rm F})}
= {\rm Tr}^{({\rm D})}\!\left\{
|\psi_{\rm out}\rangle\langle\psi_{\rm out}|
\right\} ,
\end{equation}
with
\begin{equation}
\label{2.61d}
|\psi_{\rm out}\rangle
= \prod_{\nu=1}^4 \frac{1}{\sqrt{n_\nu !}}
\left
(\sum_{\mu=1}^4 \Lambda_{\mu \nu}
\hat{\alpha}_\mu^\dagger\right)^{n_\nu} |0\rangle.
\end{equation}
We use the decomposition
\begin{equation}
\label{2.61e}
\left(
\sum_{\mu=1}^4
\Lambda_{\mu\nu}\hat{\alpha}_\mu^\dagger
\right)^{n_\nu}
= \sum_{\{k_{\nu\mu}\}}
\prod_{\mu=1}^4
\frac{n_\nu!}{k_{\nu\mu}!}
\left(
\Lambda_{\mu\nu} \hat{\alpha}_\mu^\dagger
\right)^{k_{\nu\mu}},
\end{equation}
where the (non-negative) integers $k_{\nu\mu}$ satisfy the condition
\begin{equation}
\label{2.61f}
\sum_{\mu=1}^4 k_{\nu\mu} = n_\nu \,,
\end{equation}
and arrive at
\begin{equation}
\label{2.61g}
|\psi_{\rm out}\rangle
= \sum_{\{k_\mu\}}
C_{k_1,k_2,k_3,k_4} |k_1;k_2;k_3;k_4\rangle
\end{equation}
where
\begin{eqnarray}
\label{2.61h}
\lefteqn{
C_{k_1,k_2,k_3,k_4}
}
\nonumber\\ & & \hspace{1ex} 
= \left( \prod_{\nu=1}^4 \sqrt{n_\nu!} \right)
\left( \prod_{\mu=1}^4 \sqrt{k_\mu!} \right)
\sum_{\{k_{\nu\mu}\}}
\prod_{\mu,\nu=1}^4
\frac{\Lambda_{\mu\nu}^{k_{\nu\mu}}} {k_{\nu\mu}!} \,,
\end{eqnarray}
the $k_{\nu\mu}$ satisfying the conditions
\begin{equation}
\label{2.61i}
\sum_{\mu=1}^4 k_{\nu\mu} = n_\nu\,,
\quad
\sum_{\nu=1}^4 k_{\nu\mu} = k_\mu\,.
\end{equation}
Using Eqs.~(\ref{2.61c}), (\ref{2.61g}), and (\ref{2.61h}),
the quantum state of the outgoing radiation-field modes
can easily be obtained,
\begin{equation}
\label{2.61i1}
\varrho_{\rm out}^{({\rm F})}
= \sum_{k_1,k_2}  \sum_{k'_1,k'_2}
D_{k_1,k_2,k'_1,k'_2} \, |k_1;k_2\rangle\langle k'_1;k'_2|,
\end{equation}
\begin{equation}
\label{2.61i2}
D_{k_1,k_2,k'_1,k'_2}
= \sum_{k_3,k_4} C_{k_1,k_2,k_3,k_4} C^\ast_{k'_1,k_2',k_3,k_4}\,.
\end{equation}

The density operator of the outgoing field, $\varrho_{\rm out}^{({\rm F})}$, 
can be represented in another way which more clearly shows the influence 
on the outgoing field state of the device. Let us define 
linear combinations 
\begin{equation}
\label{2.61z2}
\hat{x}_\nu^\dagger 
= \sum_{i=1}^2\Lambda_{i\nu}\hat{a}_i^\dagger 
\end{equation}
and
\begin{equation}
\label{2.61z2a}
\hat{y}_\nu^\dagger 
= \sum_{i=1}^2\Lambda_{2+i\,\nu}\hat{g}_i^\dagger 
\end{equation}
of the photonic and device operators, respectively. Making in
Eq.~(\ref{2.61d}) the insertion
\begin{equation}
\label{2.61z1}
\sum_{\mu=1}^4
\Lambda_{\mu\nu}\hat{\alpha}_\mu^\dagger 
= \hat{x}_\nu^\dagger + \hat{y}_\nu^\dagger ,
\end{equation}
Eq.~(\ref{2.61c}) then reads as
\begin{equation}
\label{2.61z3}
\varrho_{\rm out}^{({\rm F})}
= \sum_{\{p_\mu\}\{q_\nu\}} Y_{\{p_\mu\}\{q_\nu\}} 
\prod_{\mu=1}^4\hat{x}_\mu^{\dagger p_\mu}|0^{(F)}\rangle
\langle 0^{(F)}|\prod_{\nu=1}^4\hat{x}_\nu^{q_\nu} ,
\end{equation}
where 
\begin{eqnarray}
\label{2.61z4}
\lefteqn{
Y_{\{p_\mu\}\{q_\nu\}} = 
\left[\prod_{\nu=1}^4\frac{1}{\sqrt{n_\nu !}}{n_\nu \choose q_\nu}\right]
\left[\prod_{\mu=1}^4\frac{1}{\sqrt{n_\mu !}}{n_\mu \choose p_\mu}\right]
}
\nonumber\\ && \hspace{3ex} \times \,
\langle 0^{(D)}|
\left(\prod_{\nu=1}^4\hat{y}_\nu^{        n_\nu-q_\nu}\right)
\left(\prod_{\mu=1}^4\hat{y}_\mu^{\dagger n_\mu-p_\mu}\right)|0^{(D)}\rangle ,
\end{eqnarray}
and $|0^{(F)}\rangle$ ($|0^{(D)}\rangle$) is the ground state of the field 
(device). The device vacuum expectation value in Eq.~(\ref{2.61z4}) can be 
calculated by moving the
operators $\hat{y}_\nu$ from left to right and employing the commutation
relations between $\hat{y}_\nu$ and $\hat{y}_\mu^\dagger$. Then we find, 
that the coefficients $Y_{\{p_\mu\}\{q_\nu\}}$ $\!\equiv$ 
$\!Y_{\{p_\mu\}\{q_\nu\}}(Z_{\nu \mu})$ 
are functions of the matrix elements $Z_{\nu \mu}$ of the matrix 
\begin{equation}
\label{2.61z5}
{\bf Z}
=\left(\begin{array}{cc}{\bf I}-{\bf T}^+{\bf T}
&-{\bf T}^+{\bf A}\\[1ex]
-{\bf A}^+{\bf T}
&{\bf I}-{\bf A}^+{\bf A}\end{array}\right).
\end{equation}

Finally let us consider the transformation of coherent states.
\begin{eqnarray}
\label{2.62}
|\psi_{\rm in}\rangle
&=& |\gamma_1;\gamma_2,\gamma_3;\gamma_4\rangle
\nonumber\\
& = &
\prod_{\nu=1}^4 \exp\!\left(\gamma_\nu \hat{\alpha}_\nu^\dagger
-\gamma_\nu^\ast
\hat{\alpha}_\nu \right) |0\rangle .
\end{eqnarray}
   From Eq.~(\ref{2.55}) we again obtain Eq.~(\ref{2.61c}), where
$|\psi_{\rm out}\rangle$ is the coherent state
\begin{eqnarray}
\label{2.64}
|\psi_{\rm out}\rangle
&=& |\lambda_1;\lambda_2,\lambda_3;\lambda_4\rangle
\nonumber\\
& = &
\prod_{\mu=1}^4 \exp\!\left(\lambda_\mu \hat{\alpha}_\mu^\dagger
-\lambda_\mu^\ast
\hat{\alpha}_\mu \right) |0\rangle ,
\end{eqnarray}
with
\begin{eqnarray}
\lambda_\mu = \sum_{\nu=1}^4 \Lambda_{\mu\nu} \gamma_{\nu}\,.
\label{2.65}
\end{eqnarray}
   From Eqs.~(\ref{2.61c}) and (\ref{2.65}) it follows that the
outgoing modes are prepared in coherent states,
\begin{eqnarray}
\hat{\varrho}_{\rm out}^{({\rm F})}
= |\lambda_1;\lambda_2\rangle\langle\lambda_1;\lambda_2|.
\label{2.66}
\end{eqnarray}
Note that when the device is excited in a coherent state,
then the coherent amplitudes $\lambda_1$ and $\lambda_2$
of the outgoing modes are not only determined by the
characteristic transformation matrix ${\bf T}$ but also
by the absorption matrix ${\bf A}$ and the coherent-state
amplitudes of the device, as it can be seen
from Eq.~(\ref{2.65}),
\begin{eqnarray}
\label{2.68}
\lambda_i = \sum_{j=1}^2 \left( T_{ij} \gamma_j +
A_{ij} \gamma_{j+2} \right)
\end{eqnarray}
($i$ $\!=$ $\!1,2$).


\section{Summary}
\label{sum}

We have developed a quantum theory of the action of a dispersive
and absorbing optical four-port device, such as a beam splitter.
In particular we have presented formulas for calculating the 
complete quantum state of the outgoing fields from the input quantum 
state of the incoming fields and the device excitations, without any
frequency restriction. The theory is a natural extension of the 
standard theory of lossless beam splitters. According to the underlying
quantization scheme for radiation in inhomogeneous
Kramers--Kronig media, the device is described in terms of
a frequency-dependent transformation matrix that includes
transmission and reflection and a frequency-dependent absorption
matrix.  

For each frequency the action of the device has been described
in terms of a U(4) group transformation of incoming field operators and 
device operators.
Each U(4) group transformation 
can be realized in a natural way by 
the combined action of eight lossless four-port devices.  
However, each U(4) matrix 
is only determined up to a U(2) matrix. This matrix can be chosen such
that the U(4) group transformation 
is equivalent to five U(2) group transformations. That is to say,
for chosen frequency the action of an absorbing four-port device
formally corresponds to the combined action of five lossless
four-port devices.

The quantum state of the outgoing radiation
can be expected to sensitively depend on the quantum
state the device is prepared in when the incoming
fields impinge on the device. In combination
with conditional measurement this offers novel 
possibilities of quantum state manipulation. 
In particular, the theory enables one to study
the effect of resonance frequencies on quantum 
state transformation. 


\section*{acknowledgment}
We acknowledge discussions with Jan Rataj.
This work was supported by the Deutsche Forschungsgemeinschaft.


\appendix
\section{Derivation of the U(4) group matrix}
\label{U4}

Let us write the sought U(4) matrix $\mbox{\boldmath $\Lambda$}(\omega)$ as 
\begin{equation}
\label{B01}
\mbox{\boldmath $\Lambda$}(\omega)
= \left(
\begin{array}{cc}
{\bf T}(\omega) & {\bf A}(\omega)\\[1ex]
{\bf F}(\omega) &{\bf G}(\omega)
\end{array}
\right),
\end{equation}
where ${\bf T}(\omega)$ and ${\bf A}(\omega)$ are defined in 
Eqs.~(\ref{2.13}) and (\ref{2.14}) and satisfy the relation (\ref{2.16}). 
The $2\times 2$ matrices ${\bf F}(\omega)$ 
and ${\bf G}(\omega)$ are to be determined such that 
$\mbox{\boldmath $\Lambda$}(\omega)$ is unitary, i.e.,
\begin{equation}
\label{B03}
{\bf F}(\omega){\bf F}^+(\omega) 
+ {\bf G}(\omega){\bf G}^+(\omega) ={\bf I} ,
\end{equation}
\begin{equation}
\label{B04}
{\bf F}(\omega){\bf T}^+(\omega) 
+ {\bf G}(\omega){\bf A}^+(\omega) ={\bf 0}\,.
\end{equation}
   From Eq.~(\ref{B04}) we find that 
\begin{equation}
\label{B05}
{\bf F}(\omega)= 
- {\bf G}(\omega){\bf A}^+(\omega)({\bf T}^+)^{-1}(\omega).
\end{equation}
We substitute in Eq.~(\ref{B03}) for ${\bf F}(\omega)$
the result of Eq.~(\ref{B05}) and derive
\begin{equation}
\label{B06}
{\bf G}(\omega)
\left\{
{\bf I} + {\bf A}^+(\omega)
\left[{\bf T}(\omega){\bf T}^+(\omega)\right]^{-1}
{\bf A}(\omega) \right\} {\bf G}^+(\omega) = {\bf I} ,
\end{equation}
and hence
\begin{equation}
\label{B07}
{\bf I} + {\bf A}^+(\omega)
\left[{\bf T}(\omega){\bf T}^+(\omega)\right]^{-1}
{\bf A}(\omega) = \left[{\bf G}^+(\omega){\bf G}(\omega)\right]^{-1} .
\end{equation}
Recalling Eq.~(\ref{2.16}), from Eq.~(\ref{B07}) we find that
\begin{equation}
\label{B08}
{\bf G}^+(\omega){\bf G}(\omega)
= {\bf I} - {\bf A}^+(\omega){\bf A}(\omega)\,.
\end{equation}
A particular solution of Eq.~(\ref{B08}) is 
\begin{equation}
\label{B09}
{\bf G}(\omega)= {\bf C}(\omega){\bf S^{-1}}(\omega){\bf A}(\omega) ,
\end{equation}
where ${\bf C}(\omega)$ and ${\bf S}(\omega)$ are defined in
Eqs.~(\ref{2.35}) and (\ref{2.35a}), respectively. Obviously, the 
general solution reads as
\begin{equation}
\label{B11}
{\bf G}(\omega)=
{\bf D}(\omega) {\bf C}(\omega){\bf S^{-1}}(\omega){\bf A}(\omega) ,
\end{equation}
where ${\bf D}$ is an arbitrary unitary $2\times 2$ matrix. From 
Eq.~(\ref{B05}) it then follows that ${\bf F}(\omega)$ is given by
\begin{equation}
\label{B12}
{\bf F}(\omega)
= - {\bf D}(\omega) {\bf S}(\omega){\bf C}^{-1}(\omega){\bf T}(\omega) .
\end{equation}
Combining Eqs.~(\ref{B01}), (\ref{B11}), and (\ref{B12}), we obtain
\widetext
\begin{equation}
\label{B13}
\mbox{\boldmath $\Lambda$}(\omega)
=\left(
\begin{array}{cc}
{\bf T}(\omega)&{\bf A}(\omega)\\[1ex]
-{\bf D}(\omega){\bf S}(\omega){\bf C}^{-1}(\omega){\bf T}(\omega)
&{\bf D}(\omega){\bf C}(\omega){\bf S}^{-1}(\omega){\bf A}(\omega)\end{array}
\right),
\end{equation}
\narrowtext
\noindent
which reveals that for given matrices ${\bf T}(\omega)$ and 
${\bf A}(\omega)$ the U(4) matrix $\mbox{\boldmath $\Lambda$}(\omega)$ 
is only determined up to a U(2) matrix ${\bf D}(\omega)$. 


\section{Factorization of the U(4) group transformation}
\label{app}

The U(4) matrix $\mbox{\boldmath $\Lambda$}(\omega)$ in
Eq.~(\ref{B13}) can be rewritten as a product
of three U(4) matrices as follows:
\begin{eqnarray}
\label{A01a}
\mbox{\boldmath $\Lambda$}(\omega)
= \mbox{\boldmath $\Lambda$}_3(\omega) \,
\mbox{\boldmath $\Lambda$}_2(\omega) \,
\mbox{\boldmath $\Lambda$}_1(\omega) ,
\end{eqnarray}
where 
\begin{equation}
\label{A02a}
\mbox{\boldmath $\Lambda$}_1(\omega)
=\left(
\begin{array}{cc}
{\bf D}(\omega){\bf C}^{-1}(\omega){\bf T}(\omega)&{\bf 0}\\[1ex]
{\bf 0}&{\bf D}(\omega){\bf S}^{-1}(\omega){\bf A}(\omega)
\end{array}
\right),
\end{equation}
\begin{equation}
\label{A03a}
\mbox{\boldmath $\Lambda$}_2(\omega)
=\left(
\begin{array}{rr}
{\bf D}(\omega){\bf C}(\omega){\bf D}^+(\omega)
&{\bf D}(\omega){\bf S}(\omega){\bf D}^+(\omega)\\[1ex]
-{\bf D}(\omega){\bf S}(\omega){\bf D}^+(\omega)
&{\bf D}(\omega){\bf C}(\omega){\bf D}^+(\omega)
\end{array}
\right) ,
\end{equation}
\begin{equation}
\label{A03b}
\mbox{\boldmath $\Lambda$}_3(\omega)
=\left(
\begin{array}{cc}
{\bf D}^\dagger(\omega)&{\bf 0}\\[1ex]
{\bf 0}&{\bf I}
\end{array}
\right) .
\end{equation}
When we choose the matrix ${\bf D(\omega)}$ such that the matrices 
${\bf D}(\omega){\bf C}(\omega){\bf D}^+(\omega)$ and
${\bf D}(\omega){\bf S}(\omega){\bf D}^+(\omega)$ become
diagonal matrices [note that  ${\bf C(\omega)}$ and 
${\bf S(\omega)}$ defined in Eqs.~(\ref{2.35}) and (\ref{2.35a}), 
respectively, can be diagonalized by the same unitary matrix], 
then the U(4) group transformation corresponds
to five U(2) group transformations.

Let ${\bf D(\omega)}$ be the unit matrix, ${\bf D(\omega)}$ 
$\!=$ $\!{\bf I}$. In this case Eq.~(\ref{A01a}) reduces to  
\begin{eqnarray}
\label{A01}
\mbox{\boldmath $\Lambda$}(\omega)
= \mbox{\boldmath $\Lambda$}_2(\omega) \,
\mbox{\boldmath $\Lambda$}_1(\omega) ,
\end{eqnarray}
where now
\begin{equation}
\label{A02}
\mbox{\boldmath $\Lambda$}_1(\omega)
=\left(
\begin{array}{cc}
{\bf C}^{-1}(\omega){\bf T}(\omega)&{\bf 0}\\[1ex]
{\bf 0}&{\bf S}^{-1}(\omega){\bf A}(\omega)
\end{array}
\right)
\end{equation}
and
\begin{equation}
\label{A03}
\mbox{\boldmath $\Lambda$}_2(\omega)
=\left(
\begin{array}{rr}
{\bf C}(\omega)&{\bf S}(\omega)\\[1ex]
-{\bf S}(\omega)&{\bf C}(\omega)
\end{array}
\right) .
\end{equation}
The matrix $\mbox{\boldmath $\Lambda$}_2(\omega)$ can be given
by the unitary transform of a quasi-diagonal matrix
$\mbox{\boldmath $\Lambda$}_2'(\omega)$,
\begin{equation}
\label{A04}
\mbox{\boldmath $\Lambda$}_2(\omega)
=\mbox{\boldmath $\Upsilon$}^+
\mbox{\boldmath $\Lambda$}_2'(\omega)
\mbox{\boldmath $\Upsilon$} ,
\end{equation}
where
\begin{equation}
\label{A05}
\mbox{\boldmath $\Lambda$}_2'(\omega)
=\left(
\begin{array}{cc}
{\bf C}(\omega)-i{\bf S}(\omega)&{\bf 0}\\[1ex]
{\bf 0}&{\bf C}(\omega)+i{\bf S}(\omega)
\end{array}
\right)
\end{equation}
and
\begin{equation}
\label{A06}
\mbox{\boldmath $\Upsilon$}
=\frac{1}{\sqrt{2}}
\left(
\begin{array}{cc}{\bf I}&i{\bf I}\\
{i\bf I}&{\bf I}
\end{array}\right).
\end{equation}
Combining Eqs.~(\ref{A01}) and (\ref{A04}), we obtain
\begin{equation}
\label{A06a}
\mbox{\boldmath $\Lambda$}(\omega)
=\mbox{\boldmath $\Upsilon$}^+
\mbox{\boldmath $\Lambda$}_2'(\omega)
\mbox{\boldmath $\Upsilon$} 
\mbox{\boldmath $\Lambda$}_1(\omega),
\end{equation}
which corresponds to a decomposition of the U(4) group transformation
into eight U(2) group transformations. 

Using Eqs.~(\ref{2.28}) and (\ref{A01}) and
recalling Eqs.~(\ref{2.37}) -- (\ref{2.39}), we may write
\begin{eqnarray}
\label{A07} 
\hat{\mbox{\boldmath $\beta$}}(\omega) &=&  
\mbox{\boldmath $\Lambda$}_2(\omega) \,
\mbox{\boldmath $\Lambda$}_1(\omega) \,
\hat{\mbox{\boldmath $\alpha$}}(\omega)
\nonumber\\
& =&
\mbox{\boldmath $\Lambda$}_2(\omega)\,
\hat{U}_1^\dagger
\hat{\mbox{\boldmath $\alpha$}}(\omega)
\hat{U}_1
\nonumber\\
&=&\hat{U}_1^\dagger \,
\mbox{\boldmath $\Lambda$}_2(\omega)\,
\hat{\mbox{\boldmath $\alpha$}}(\omega)\,
\hat{U}_1
\nonumber\\
&=&\hat{U}_1^\dagger\,\hat{U}_2^\dagger\,
\hat{\mbox{\boldmath $\alpha$}}(\omega)\,
\hat{U}_2\hat{U}_1
\, = \, \hat{U}^\dagger\,
\hat{\mbox{\boldmath $\alpha$}}(\omega)\,
\hat{U},
\end{eqnarray}
with
\begin{equation}
\label{A07a}
\hat{U} \equiv
\hat{U}[\mbox{\boldmath $\Lambda$};\hat{\mbox{\boldmath $\alpha$}}]
= \hat{U}[\mbox{\boldmath $\Lambda$}_2;\hat{\mbox{\boldmath $\alpha$}}]
\hat{U}[\mbox{\boldmath $\Lambda$}_1;\hat{\mbox{\boldmath $\alpha$}}].
\end{equation}
Here, $\hat{U}_i$ $\!\equiv$ $\!\hat{U}[\mbox{\boldmath $\Lambda$}_i;
\hat{\mbox{\boldmath $\alpha$}}]$ ($i$ $\!=$ $\!1,2$)
is given by Eq.~(\ref{2.38}), 
with $\mbox{\boldmath $\Phi$}_i(\omega)$ in place of
$\mbox{\boldmath $\Phi$}(\omega)$, and 
\begin{equation}
\label{A07b}
\exp[-i\mbox{\boldmath $\Phi$}_i(\omega)] = 
\mbox{\boldmath $\Lambda$}_i(\omega).
\end{equation}
   From the quasi-diagonal structure of
$\mbox{\boldmath $\Lambda$}_1(\omega)$, Eq.~(\ref{A02}), 
it then follows that
\begin{equation}
\label{A08}
\mbox{\boldmath $\Phi$}_1(\omega)
= \left(
\begin{array}{cc}
{\bf W}_1(\omega)&{\bf 0}\\[.5ex]
{\bf 0}&{\bf 0}
\end{array}
\right) + \left(
\begin{array}{cc}
{\bf 0}&{\bf 0}\\[.5ex]
{\bf 0}&{\bf W}_2(\omega)\end{array}\right) ,
\end{equation}
where
\begin{equation}
\label{A09}
\exp\!\left[-i{\bf W}_1(\omega)\right]
= {\bf C}^{-1}(\omega){\bf T}(\omega)
\end{equation}
and
\begin{equation}
\label{A10}
\exp\!\left[-i{\bf W}_2(\omega)\right]
= {\bf S}^{-1}(\omega){\bf A}(\omega) .
\end{equation}
Thus, $\hat{U}[\mbox{\boldmath $\Lambda$}_1;\hat{\mbox{\boldmath $\alpha$}}]$
can be expressed in terms of two unitary operators of the type given in
Eq.~(\ref{2.44}) [together with Eq.~(\ref{2.44a})],
\begin{equation}
\label{A10a}
\hat{U}[\mbox{\boldmath $\Lambda$}_1;\hat{\mbox{\boldmath $\alpha$}}]
= \hat{U}[{\bf S}^{-1}{\bf T};\hat{\bf g}]
\,\hat{U}[{\bf C}^{-1}{\bf T};\hat{\bf a}] .
\end{equation}

To decompose
$\hat{U}[\mbox{\boldmath $\Lambda$}_2;\hat{\mbox{\boldmath $\alpha$}}]$,
we note that Eq.~(\ref{A04}) implies that
\begin{equation}
\label{A11}
\hat{U}[\mbox{\boldmath $\Lambda$}_2;\hat{\mbox{\boldmath $\alpha$}}]
= \hat{U}[\mbox{\boldmath $\Lambda$}_2';
\mbox{\boldmath $\Upsilon$}(\omega)
\hat{\mbox{\boldmath $\alpha$}}] ,
\end{equation}
where
\begin{equation}
\exp\!\left[-i \mbox{\boldmath $\Phi$}_2'(\omega) \right] = 
\mbox{\boldmath $\Lambda$}_2'(\omega) 
\label{A12}
\end{equation}
and
\begin{equation}
\mbox{\boldmath $\Upsilon$}(\omega)
\hat{\mbox{\boldmath $\alpha$}}(\omega)
= \frac{1}{\sqrt{2}}
\left(
\begin{array}{ccc}
\hat{\bf a}(\omega)\!\! &+&\!\! i \hat{\bf g}(\omega) \\[.5ex]
i\hat{\bf a}(\omega)\!\! &+&\!\! \hat{\bf g}(\omega)
\end{array}
\right).
\label{A16}
\end{equation}
The quasi-diagonal structure of
$\mbox{\boldmath $\Lambda$}_2'(\omega)$, Eq.~(\ref{A05}),
enables us to write
\begin{equation}
\label{A13}
\mbox{\boldmath $\Phi$}_2'(\omega)
= \left(
\begin{array}{cc}
{\bf W}_3(\omega)&{\bf 0}\\[.5ex]
{\bf 0}&{\bf 0}
\end{array}
\right) + \left(
\begin{array}{cc}
{\bf 0}&{\bf 0}\\[.5ex]
{\bf 0}&-{\bf W}_3(\omega)
\end{array}
\right),
\end{equation}
where
\begin{equation}
\label{A14}
\exp\!\left[-i{\bf W}_3(\omega)\right]
= {\bf C}(\omega)-i{\bf S}(\omega) ,
\end{equation}
so that
$\hat{U}[\mbox{\boldmath $\Lambda$}_2;\hat{\mbox{\boldmath $\alpha$}}]$
can also be expressed in terms of two unitary operators of the type
given in Eq.~(\ref{2.44}) [together with Eq.~(\ref{2.44a})],
\begin{eqnarray}
\label{A17}
\lefteqn{
\hat{U}[\mbox{\boldmath $\Lambda$}_2;\hat{\mbox{\boldmath $\alpha$}}]
= \hat{U}[{\bf C}\!+\!i{\bf S};(i\hat{\bf a}\!+\!\hat{\bf g})/\sqrt{2}]
}
\nonumber\\&&\hspace{15ex}\times\,
\hat{U}[{\bf C}\!-\!i{\bf S};(\hat{\bf a}\!+\!i\hat{\bf g})/\sqrt{2}].
\end{eqnarray}
Recalling the definitions of $\hat{\bf d}_j(\omega)$, Eq.~(\ref{2.45b}), and
${\bf P}$, Eq.~(\ref{2.45c}), it is seen that
$(\hat{\bf a}$ $\!+$ $\!i\hat{\bf g})/\sqrt{2}$
and $(i\hat{\bf a}$ $\!+$ $\!\hat{\bf g})/\sqrt{2}$
can be given by
\begin{eqnarray}
\label{A17a}
\lefteqn{
\frac{1}{\sqrt{2}}
\left(
\begin{array}{ccc}
\hat{a}_j(\omega)\! \!&+& \!\!i \hat{g}_j(\omega) \\[.5ex]
i\hat{a}_j(\omega) \!\!&+& \!\!\hat{g}_j(\omega)
\end{array}
\right)
= {\bf P}\,\hat{\bf d}_j(\omega) 
}
\nonumber \\[1ex] && \hspace{15ex}
 =  \hat{U}^\dagger[{\bf P};\hat{\bf d}_j]\,
\hat{\bf d}_j(\omega) \, \hat{U}[{\bf P};\hat{\bf d}_j] 
\end{eqnarray}
($j$ $\!=$ $\!1,2$), and hence
\begin{eqnarray}
\label{A17b}
\lefteqn{
\hat{U}[{\bf C}\!-\!i{\bf S};(\hat{\bf a}\!+\!i\hat{\bf g})/\sqrt{2}]
= \hat{U}^\dagger[{\bf P};\hat{\bf d}_1]\,
\hat{U}^\dagger[{\bf P};\hat{\bf d}_2]\,
}
\nonumber\\ && \hspace{15ex} \times \,
\hat{U}[{\bf C}\!-\!i{\bf S};\hat{\bf a}]
\,\hat{U}[{\bf P};\hat{\bf d}_2] 
\,\hat{U}[{\bf P};\hat{\bf d}_1],
\end{eqnarray}
\begin{eqnarray}
\label{A17c}
\lefteqn{
\hat{U}[{\bf C}\!+\!i{\bf S};(i\hat{\bf a}\!+\!\hat{\bf g})/\sqrt{2}]
= \hat{U}^\dagger[{\bf P};\hat{\bf d}_1]\,
\hat{U}^\dagger[{\bf P};\hat{\bf d}_2]\,
}
\nonumber\\ && \hspace{15ex} \times \,
\hat{U}[{\bf C}\!-\!i{\bf S};\hat{\bf g}]
\,\hat{U}[{\bf P};\hat{\bf d}_2] 
\,\hat{U}[{\bf P};\hat{\bf d}_1].
\end{eqnarray}

\end{document}